\documentstyle[12pt,aaspp4]{article}

\def\ni{\noindent}
\def\.{\mathaccent 95}
\def\beq{\begin{equation}}
\def\ee{\end{equation}}

\def\be{\beta}

\def\frac#1#2{{\textstyle{{#1}\over {#2}}}}
\def\ni{\noindent}
\def\lsim{\mathrel{\rlap{\lower4pt\hbox{\hskip1pt$\sim$}}
    \raise1pt\hbox{$<$}}}
\def\gsim{\mathrel{\rlap{\lower4pt\hbox{\hskip1pt$\sim$}}
    \raise1pt\hbox{$>$}}}
\def\sqr#1#2{{\vcenter{\vbox{\hrule height.#2pt
         \hbox{\vrule width.#2pt height#1pt \kern#1pt
         \vrule width.#2pt}
         \hrule height.#2pt}}}}

\newbox\grsign \setbox\grsign=\hbox{$>$} \newdimen\grdimen
\grdimen=\ht\grsign
\newbox\simlessbox \newbox\simgreatbox
\setbox\simgreatbox=\hbox{\raise.5ex\hbox{$>$}\llap
     {\lower.5ex\hbox{$\sim$}}}\ht1=\grdimen\dp1=0pt
\setbox\simlessbox=\hbox{\raise.5ex\hbox{$<$}\llap
     {\lower.5ex\hbox{$\sim$}}}\ht2=\grdimen\dp2=0pt

\def\x      {{\hbox{X-ray}}}

\def\kms    {~km~s$^{-1}$}

\def\etal   {{\it et~al.}}

\def\doublespace {\smallskipamount=6pt plus2pt minus2pt
                  \medskipamount=12pt plus4pt minus4pt
                  \bigskipamount=24pt plus8pt minus8pt
                  \normalbaselineskip=24pt plus0pt minus0pt
                  \normallineskip=2pt
                  \normallineskiplimit=0pt
                  \jot=6pt
                  {\def\smallskip {\vskip\smallskipamount}}
                  {\def\medskip   {\vskip\medskipamount}}
                  {\def\bigskip   {\vskip\bigskipamount}}
                  {\setbox\strutbox=\hbox{\vrule 
                    height17.0pt depth7.0pt width 0pt}}
                  \parskip 12.0pt
                  \normalbaselines}

\font\gkvec=cmmib10                         
\def\bomega{\hbox{{\gkvec\char33}}}                  

\def\lb{\langle}
\def\rb{\rangle}

\def\bw{\overline {\omega}}
\def\bv{\overline V}

\def\ts{\times}
\def\lb{\langle}
\def\rb{\rangle}

\def\bfv{{\bf v}}

\def\bfj{{\bf j}}

\def\bfw{{\bomega}}

\def\bfb{{\bf b}}

\def\b0{b^{(0)}}
\def\v0{v^{(0)}}
\def\w0{\omega^{(0)}}
\def\bb0{\bfb^{(0)}}
\def\bv0{\bfv^{(0)}}
\def\bw0{\bfw^{(0)}}
\def\bj0{\bfj^{(0)}}

\def\ni{\noindent}

\def\ea{{\it et al.}}
\def \msols{\rm{M}$_\odot$~yr$^{-1}$}

\begin{document}

\def\be{\begin{equation}}
\def\ee{\end{equation}}
\def\lab#1{\label{#1}}
\def\lrp#1{\left(#1\right)}
\def\BV{{\bf V}}
\def\etal{{\it et al.}\ }
\def\OV{\overline{\bf V}}
\def\E{{\bf E}}
\def\x{{\bf x}}

\def\pref#1{(\ref{#1})}
\def\beq{\begin{eqnarray}}
\def\eeq{\end{eqnarray}}
\def\nn{\nonumber}
\def\nt{\nabla\times}
\def\OE{\overline{\bf E}}
\def\A{{\bf A}}
\def\lra#1{\left\langle #1\right\rangle}
\def\bv{\bf v}
\def\OB{\overline{\bf B}}
\def\ove{\overline{E}}
\def\cnt{\cdot\nabla\times}
\def\b{{\bf b}}
\def\ob{\overline{B}}
\def\ao{\alpha\hbox{-}\Omega}
\def\B{{\bf B}}

\centerline {\bf MHD Stellar and Disk Winds: Application to 
Planetary Nebulae}
\medskip
\centerline{Eric G. Blackman$^{1,2}$, Adam Frank$^1$, and Carl Welch$^1$}  
\centerline{1. Department of Physics \& Astronomy, University of Rochester, Rochester, NY 14627}
\centerline{2. ITP, University of California, Santa Barbara, CA 93106}
\medskip
\centerline{(accepted to ApJ)}
\medskip 

\bigskip 
\medskip
\centerline{\bf Abstract}
\medbreak

MHD winds can emanate from both stars and surrounding disks.  
When the two systems are coupled by accretion, it is of interest to
know how much wind power is available and 
which (if either) of the two rotators dominates that power.
We investigate this in the context of multi-polar
planetary nebulae (PNe) and proto-planetary nebulae (PPNe),
for which recent observations have revealed the need for  
a wind power source in excess of that available from radiation driving,
and a possible need for magnetic shaping.  
We calculate the MHD wind power from 
a coupled disk and star, where the former results from binary
disruption. The resulting wind powers  
depend only on the accretion rate and stellar properties.
We find that if the stellar envelope  were initially slowly
rotating, the disk wind would dominate throughout the evolution.
If the envelope of the star were rapidly  rotating, the stellar wind
could initially be of comparable power to the disk wind
until the stellar wind carries away the star's angular momentum.  
Since an initially rapidly rotating star can have its spin and
magnetic axes misaligned to the disk, multi-polar outflows can result
from this disk wind system.  For times greater than a spin-down time,
the post-AGB stellar wind is slaved to the disk for both slow and rapid
initial spin cases and the disk wind luminosity dominates.
We find a reasonably large parameter space where a hybrid star+disk MHD
driven wind is plausible and where both or either
can account for PPNe and PNe powers.
We also speculate on the morphologies which may emerge from the coupled system.
The coupled winds might help explain the shapes of a number of
remarkable multi-shell or multi-polar nebulae. 
Magnetic activity such as X-ray flares may
be associated with the both central star and the disk and would 
be a valuable diagnostic for the dynamical role of MHD processes in PNe.
\bigskip

\ni {\bf Subject headings}: jets and outflows; accretion, accretion
disks; 
planetary nebulae: general; ISM: magnetic fields: MHD; stars: AGB and
post-AGB


\vfill
\eject

\section{Introduction}

Planetary Nebulae (PNe) and proto-Planetary Nebulae (PPNe) are believed
to be the penultimate evolutionary stage of low and intermediate mass
stars 
(zero age main sequence mass range 
$0.8 M_\odot < M_{zams} \le 5 M_\odot$). PNe appear on the sky as expanding 
ionized
plasma clouds surrounding a hot central star.  As the resolution
of ground based telescopes increased, the ``typical'' shape of a PN
went from spherical (Osterbrock 1972) to elliptical and/or bipolar
(Balick 1987).  More recently, deeper and higher resolution studies
(particularly those with the HST) have shown many PNe with narrow
collimated features that are better described as {\it jets} than
{\it bipolar lobes}.  In some cases the jets appear as extremely well
collimated bipolar outflows: M2-9 (Balick 2000): Hen-401 (Sahai 2000),
AFGL 2688 (The Egg) (Sahai \ea 1998).  In other cases they appear as
FLIERS (Fast Low-Ionization Emission Regions), a single knot expanding
away from a (usually) elliptical nebula. In addition to the presence
of jets it also appears that many PNe show evidence for multi-shell
structures or multi-polarity. In these objects two nested pairs
of bipolar lobes appear.  In multi-shell structures the lobes are
aligned along the same axis of symmetry (Hb 12: Welch \ea 1999)
while multi-polar cases show different orientations for the inner
and outer lobes (M2-46: Manchado \ea 2000).

While considerable progress has been made in understanding  the
hydrodynamics or magneto-hydrodynamics of shaping bipolar lobes
(see Frank 1999 for a review), the origin of jets, point
symmetry and multi-polar outflows in PNe poses fundamental challenges
for theory.

The basic hydrodynamics of bipolar outflows can be
described by the classic Generalized Interacting Stellar
Winds (GISW) model (Kwok \ea 1978, Kahn \& West 1986,
Balick 1987, Icke 1988).  In this scenario a slow ($10$
\kms), dense ($10^{-4}$ \msols) toroidal wind expelled during the AGB
is followed by a fast ($1000$ \kms), tenuous ($10^{-7}$ \msols ) wind
driven off the contracting proto-white dwarf during the PNe phase.
The GISW model can explain bipolar morphologies and even jets.
Point-symmetry and multi-polar bubbles however, do not fall easily into
line with the concept of a large-scale collimating torus.

Models invoking a toroidal magnetic field embedded in a normal
radiation driven stellar wind have recently shown some promise.
This so-called {\it Magnetized Wind Bubble (MWB)} model was first
proposed by (Chevalier \& Luo 1994) and has been studied numerically
by (Rozyczka \& Franco 1996) and (Garc\'ia-Segura \ea 1999).  In these
models the field at the star is dipolar but assumes a toroidal topology
due to rapid stellar rotation.  Collimation is not activated until
the wind passes through the inner shock.  Then hoop stresses associated
with the toroidal field dominate over isotropic gas pressure forces
and material is drawn towards the axis producing a collimated flow.
This mechanism has been shown capable of producing a wide variety of
outflow morphologies including well collimated jets.  When precession
of the magnetic axis is included in fully 3-D simulations, the MWB
model is capable of recovering point-symmetric morphologies as well
(Garc\'ia-Segura 1997).  
It is noteworthy that the wind in these models differs substantially from 
standard MHD driven winds in disk and stellar wind contexts, and the 
fundamental issue of collimation before the wind-wind interaction remains to be
addressed.  In addition it is not clear that such models can be applied
to multi-polar bubbles due to the difference in collimation orientations.

Finally neither the GISW or MWB models fully address the origin of
the wind. This becomes a critical issue when one is dealing with
Proto-Planetary Nebulae (PPNe).  There is growing evidence suggesting that the PN
shaping process begins in the PPN phase via well collimated jets
(Sahai \& Trauger 1998).  The creation of fast ($V > 100$ km/s)
high mass loss rate ($\dot{M} > 10^{-6}$ \msols) winds during  these
early phases when the star is an F or G type is not easily explained
via classic line driven wind theory 
as the available power is insufficient (Alcolea et al. 2000).  
Thus for PPN, theorists face
a dilemma similar to the problem of YSO outflows in that they must
explain {\it both} wind acceleration and collimation (Pudritz 1991, 
K\"onigl \&  Ruden 1993, Shu \ea 1994).

The above considerations motivate further study of magnetized
wind models for the launching and collimation of outflows from PN.
In particular, the basic question of the available MHD wind
power is important to address.

Magnetically driven outflows are a leading 
paradigm for a  ubiquity of source classes 
in nature. This includes such contexts as protostellar jets
(c.f. Smith 1998; Frank 1999 for reviews), the Solar wind (c.f. Parker
1979), 
jets of Galactic accretors (c.f. Mirabel and Rodriguez 1999), 
active galactic nuclei (c.f. Ferrari 1998; Blandford 2000 for reviews), 
and gamma-ray bursts (e.g. Usov 1992; Duncan \& Thomspon 1992;
Blackman et al. 1996; 
Meszaros \& Rees 1997; Ruderman et al. 2000
).  While the parameter regimes are varied 
and the extent and nature of collimation, particle acceleration,
and radiation process are different in the different settings, an
underlying principle remains:  magnetic fields can act as a drive
belt between gravity and energy deposition at large radii, extracting
the rotational energy of the rotator into an outflow. We note that
the potential for accretion disks to exist in PPN systems has been
raised  (Morris 1987; Soker \& Livio 1994; Reyes-Ruiz \& Lopez 1999).

In the case of PN, as in other star-disk or compact object-disk
systems, the magnetically driven outflows can in principle emanate
from both the central object or the disk. 
Because the relative collimation and power evolution of the 
disk and stellar winds can be different, it may not be unreasonable
to see outflows of different character coexisting in the same object.
In order to investigate this properly however, it is necessary to allow
that the disk and star are coupled and this requires a model
for the disk evolution.

A careful investigation of the relevant disk formation 
is presented in Reyes-Ruiz \& Lopez (1999).  The disk
in this study forms when a binary system undergoes common envelope
evolution.  The ejection of the envelope involves transfer of angular
momentum from the secondary to the envelope after which the secondary
loses enough angular momentum to fall to a separation  such that it can fill
its Roche lobe and form the disk.  Reyes-Ruiz \& Lopez (1999) 
discuss a number of important
constraints on the constituents and properties 
of binaries which lead to disk
formation.  The most likely system turns out to be an evolved AGB star
with mass $2.5< M_{agb}/M_\odot < 5$ for the primary, with a secondary
of
mass $\lsim 0.08M_\odot$ and with an initial binary separation of $>
50-100 R_\odot$ (Iben 1991).  The AGB star will shed 80\% of its mass
during the common envelope ejection, leaving a post AGB stellar core
of Mass $M_c$ surrounded by a thin residual convective shell.

The time scale for the disk to move to its inner inner radius is of
order or shorter than the viscous time scale.  Even for systems whose
initial  outer radii are of order $100R_{\odot}$, the relevant time
scale would be no more than a few years.  Since the disk would form
only after the period of common envelope ejection, $<$ 1yr, we would
expect the disk and the stellar shell to form
concurrently within a time of order a few years.
Because the relevant stellar field will be exposed once the envelope
ejection occurs, and because any stellar dynamo growth time in the
disk (discussed later) is less than 1 year, we suggest the rotating
stellar shell and disk could both support co-operative jets
at a time coincident within about a year after the disk forms.  This 
represents the initial time appearing in our calculations.

In section 2, we determine the MHD wind power from PPN stars and disks.
This requires a calculation of the expected field strength in the disk and
a model of how the accretion affects the stellar field, the
stellar spin, and the disk inner radius.
We study the cases for which the stellar shell is initially
rapidly or slowly rotating, and calculate the  maximum
disk and stellar magnetic wind powers for each case. In section 3 we
discuss the observational implications of the results
of the power calculation and speculate on the magnetic shaping.
We summarize and conclude in section 4.

\section{Winds from Disks and Stars}

There are many models of magnetic jet production and collimation in
the literature.  The extent and explicit details of collimation are
perhaps less agreed upon than the principles of launching.  
Magnetic launching
mechanisms can be divided into two basic classes, ``spring'' mechanisms
(Uchida \& Shibata 1985; 
Contopoulos 1995 Lynden-Bell 1996) and ``fling'' mechanisms (Blandford
\& Payne 1982; Lovelace et al. 1987). 
In the former class of models, the initial driving force is a
magnetic spring, i.e.  the magnetic field energy density 
is of order the kinetic energy density in the 
disk and the outflow is driven by toroidal field
pressure.  In the fling mechanism, 
the initial driving is centrifugal along poloidal field lines.
Higher up in the corona the wind becomes magnetically driven.  
In this study we compute the total available magnetic luminosity $L_m$
from the disk and central star. 
The magnetic luminosity provides an upper limit to the associated wind
kinetic luminosity. The $L_m$ is the integral of the Poynting flux over the
surface area of the rotator. We thus have  
\beq
L_w \sim  {\dot M}_w V_{w}^2 \sim 
\epsilon L_m = \epsilon \int ({\bf E} \times {\bf B}) \cdot d{\bf S}
\sim \epsilon \int_{R_{i}}^{R_{o}} (\Omega R) B_pB_\phi RdR, 
\eeq
where $\epsilon$ is an efficiency, and $B_\phi$ and $B_p$ are the 
toroidal and poloidal field respectively, $\Omega$ and $R$ are the 
rotational frequency and radius of the magnetized wind source, 
${\dot M}_m$ is the wind outflow rate, and $V_{w}$ is the wind outflow
speed.  In what follows we discuss this equation in the context of
PPN disks and post AGB stars. For all cases of interest, for disks 
the integrand falls off fast enough such that the inner radius
matters most.  Thus for disks we have 
\beq
L_{dw}
\sim \epsilon_d L_{dm}\sim \epsilon_d B_p(R_i)B_\phi(R_i)\Omega_{d}(R_i) R_i^3,
\label{1a}
\eeq
where $\Omega_d$ is the disk angular speed $L_{dm}$ is the available magnetic
luminosity associated with the disk, and $\epsilon_d$ is the
unknown efficiency factor.
For stars, we have
\beq
L_{sw}
\sim \epsilon_{s}L_{sm}\sim B_p(R_*)B_\phi(R_*)\Omega_{*}(R_*) R_*^3,
\label{1b}
\eeq
where $\Omega_*$ is the angular speed of the stellar surface in which
the field lines are anchored, $L_{sm}$ is the available  
magnetic luminosity associated with the star, 
and $\epsilon_s$ is the stellar wind 
efficiency factor. The above disk and stellar 
wind luminosities are not independent.
The poloidal and  toroidal fields
are coupled through differential rotation in both the disk
and star, and the angular velocity of the star may depend
on the accretion.  In addition, the inner radius of the 
disk depends on the stellar magnetic field.  
Thus  (\ref{1a}) and (\ref{1b}) are very much interdependent.

We now proceed to solve for the coupling relations and consider
the stellar and disk MHD wind luminosities for two cases:
\begin{itemize}
\item{The star is initially a strongly magnetized rotator independent
of the accretion.}
\item{The star is initially a slow rotator but is spun up by the disk
accretion.}
\end{itemize} 
As we will see, the evolution in both cases depends on
estimates of the disk and star magnetic fields. We must also
solve for the disk inner radius, and the  
stellar surface angular speed evolution.

\subsection{Magnetic Field of Disk}

If the jet/wind is due to a magnetic field in the disk, this field must
be produced in situ for the PN case.  This is because the companion
star, the progenitor of the disk,  would have to be a brown dwarf or a
Jupiter type planet (Reyes-Ruiz \& Lopez 1999)
which typically would only have fields of order or
less than 10G.  Even a field several orders of magnitude larger  would
be far too small to have any influence on the disk, which forms from
shredding this mass, if the disk field were only that resulting from
flux freezing.  We estimate the field strength roughly from dynamo or
turbulent amplification in the disk.  An alternative dynamo approach
to ours is given by Reyes-Ruiz \& Stepinski (1995).

For an accretion disk whose angular momentum is transported by
a magneto-shearing instability (c.f. Balbus-Hawley 1991, 1998),
the time scale for growth of unstable modes and the time scale for
the largest eddy turnover time are both approximately equal to the
rotation time scale.  Very roughly,  the eddy turnover time can be
written
as $L/v_T$ where $v_T$ is the dominant turbulent velocity and $L$ is
the dominant correlation scale. 
We can estimate the strength of the mean magnetic field
by first estimating the total magnetic energy.  Using the
Shakura-Sunyaev (1973) viscosity prescription
\be
\nu=\alpha_{ss} c_sH\sim v_T^2/\Omega,
\label{ss}
\ee
where $\alpha_{ss}, c_s, H$ and $L/v_T$ are the disk viscosity
parameter,
the sound speed, the scale height and the dominant eddy turnover time
respectively.
Using the fact that turbulent stretching leads to $v_A\sim v_T$,
where $v_A$ is the Alfv\'en speed (ignoring 
a possible factor of $\sqrt 2$ on $v_T$, e.g. Balbus \& Hawley 1998) , 
and $\Omega R =c_s R/H$ for a thin accretion disk, we then have
straight away 
\be
v_A^2=\alpha_{ss} c_s^2.
\label{ss2}
\ee
When modeled as a mean-field  $\alpha_{d}-\Omega$ dynamo (c.f. Parker
1979;
Reyes-Ruiz \& Stepinski 1995; 
Blackman 2000), it can be shown that due to the differential shear,
the mean toroidal field exceeds the saturated mean poloidal field
strength in the disk by a factor of $H/L$. 
From (\ref{ss}) and (\ref{ss2}) we have $L\sim \alpha_{ss}^{1/2}H$
and thus 
\beq
B_p \sim \alpha_{ss}^{1/2}B_\phi.
\label{ss2b}
\eeq
Whether this relation holds true in the base of the disk corona
where the jet would launch is unclear, but
since the field in the corona is no greater than the field
in the disk, the overall disk mean field provides a good upper limit
at least to the total mean field energy density.
Note that the total mean stress $\lb B_\phi B_p \rb
=\lb {\overline B}_\phi{\overline B}_p\rb
+\lb b_\phi b_p\rb$, where the second term is a correlation
of fluctuating fields. Actually, both of these terms can drive a wind.
Exploring this point is outside the scope of the present paper, and here 
we  simply stick with the standard approach 
(e.g. Blandford \& Payne 1982) where the 
large scale fields drive the wind.  Thus we ignore the 
$\lb b_\phi b_p\rb$ term.

The upper limit of the mean field is of order the random field, so
from (\ref{ss2}) we have, for the dominant disk field component,  
\beq
{ B}_\phi^2
\sim 4\pi{\rho_d}\alpha_{ss} c_s^2.
\label{ss3}
\eeq
From the mass continuity equation of the disk we have,
\beq
\rho_{d}={\dot M}_a/(4v_R\pi H R),
\label{ss4}
\eeq
where $v_R$
is the radial infall speed.
For (thin or ADAF) disks 
\beq
v_R \sim \alpha_{ss} c_sH/R\sim\alpha_{ss} v_k(H/R)^2, 
\label{ss4a}
\eeq
where $v_k^2=GM/R$, the Keplerian speed.
Then using 
(\ref{ss3}) and (\ref{ss4}) 
and (\ref{ss4a}), 
we have 
\beq
{B}_\phi{ B}_p=\alpha_{ss}^{1/2}{B}_\phi^2 
={\alpha_{ss}^{1/2}(R/H)v_k(R) {\dot M}_a\over R^2}.
\label{ss5}
\eeq
This expression can be used to estimate either the magnetic
field strength or Maxwell stress avialable in the disk
for launching a wind.

\subsection{Magnetic field of the stellar shell}

The relevant stellar field is the field at the  
anchoring radius of the post AGB star from which the wind emanates.
Observational interpretation suggests that 
the common envelope ejection, which would precede disk formation, 
removes $\sim 80$\% of an initially $\sim 3M_\odot$ 
star (Sch\"onberner 1993). Convective stellar models for a 3$M_{\odot}$
star
(Kawalar, private communication 2000) 
then tell us that outer layer will have a convective 
shell containing a mass of order $0.01M_\odot$. 
We allow for  a distinction between
the core of the star $M_c$ and the mass in which
the field is anchored, $M_*$.
The latter is likely to be equal to the core mass if the field is generated
by an AGB interface dynamo (Blackman et al. 2000), 
but if the field were somehow anchored
only in the thin shell, $M_*$ might be $<< M_c$.

Since the convective shell is likely to be differentially
rotating with respect to the core 
we assume that the shear rate is of order
the anchoring material's angular speed, $\Omega_{*}$. 
The shear will stretch a dynamo produced poloidal
field linearly in time.  The linear streching could operate coherently
over a vertical diffusion time $\tau_D\sim R_*^2/\eta_T$,
where $\eta_T\sim Lv_T$ is the turbulent diffision coefficient
for turbulent velocity $v_T$ and scale $L$. 
(This would need to be calculated for specific 
interface dynamo models, c.f. Parker 1993; 
Markiel \& Thomas 1999; Blackman et al 2000).
Using numbers from Kawaler (private communication 2000),
the diffusion time could be 
of order a few times the convective overturn time for the largest
eddies in the post-AGB star, about $0.05$yr.
In any case, for linear growth of the toroidal field,
the toroidal and poloidal fields are then related by 
\beq
B_{\phi*}\sim B_{p*}\Omega_{*}\tau_D,
\label{fields}
\eeq
which follows from the magnetic induction equation.
Although the poloidal field $B_{p*}$ will likely also depend on 
$\Omega_{*}$, it will also depend on other properties of the
turbulence and the mechanism of field origin 
(c.f. Parker 1979, 1993; 
Pascoli 1997; Soker 1998; Markiel \& Thomas 1999; Blackman et al. 2000). 
For an initial treatment of the problem, we simply 
consider $B_{p*}$ as an input condition. 
\subsection{Disk inner radius}
Now that we have the stellar magnetic field, we can calculate the disk
inner
radius.  Related calculations were performed by Ghosh \& Lamb (1978)
and Ostriker \& Shu (1995).  
We estimate the inner radius
by balancing the infall ram pressure from the disk with the 
magnetic pressure of the star. For the infall velocity we use the
free fall velocity, as that corresponds to the field strength 
at which the disk must absolutely truncate.  Thus our calculation
represents a lower limit on the inner radius.  Note for example
that when the Alfv\'en speed in the disk, using the stellar
field strength, equals the sound speed there, magneto-shearing
instabilities will be shut off.  At this field strength
the nature of the angular momentum transport must change
(to e.g. global modes, B.Chandran, E.Ostriker, 2000 personal communication). 
We assume that a suitable change occurs  
and thus still use the larger Keplerian speed for the balance below.

Using the $r^{-3}$ dependence for
a dipole field and considering only the values at the equator $r=R$, we
have
\beq
{1 \over 4\pi}B_*^2\left({R_* \over R_i}\right)^6 \simeq 
\rho_{d,i} v_{k,i}^2,
\label{ss6a}
\eeq
where $v_{k,i}$ is the Keplerian ($\sim$ free-fall) speed 
at $R_i$. Note that 
when $\Omega_*\tau_D\gsim 1$, $B_{*}^2\sim B_{\phi*}^2$,
otherwise $B_{*}^2\sim B_{p*}^2$. This is important because
the angular speed couples in the former case but not in the latter.

The next step is to use the 
mass continuity equation (\ref{ss4}) for $R_i$.
The result is
\begin{eqnarray}
R_{i}  =  B_{p*}^{4/7}(1+\Omega_{*}^2\tau_D^2)^{2/7}
\alpha_{ss}^{2/7}(H_i/R_i)^{6/7}
R_*^{12/7}(GM_c)^{-1/7}{\dot M}_a^{-2/7}. 
\label{ss7a}
\end{eqnarray}
The inner radius will be time dependent because of the
time dependent accretion rate and angular speed of the anchoring material.  
We now compute the evolution of this angular speed.

\subsection{Angular Speed of the Star}

The disk and stellar wind luminosities depend on the angular
velocity $\Omega_*$ of the stellar material where the field is 
anchored.  As shown above, 
$\Omega_*$ is one factor which determines the magnetic pressure
of the star, and thus $R_i$.  
For small $\tau_D$, $\Omega_*$ decouples from $R_i$.
 
If material is accreting onto the star from the disk at
rates based on the disk formation models of Reyes-Ruiz and
Lopez (1999), then 
\beq
{\dot M}_a\sim 6\ts 10^{22}\left(t \over {\rm 1yr}\right)^{-5/4}{\rm g/s},
\label{rrl1}
\eeq
and there may be some contribution to the rotation
from the accreted material.
There are many complications in this accretion process.
Here we simply assume that the material imparts its angular momentum
to the shell (e.g. via the magnetic field) as it accretes.
We also assume that the accreted material 
does not change the thickness of the star.  
The rotation speed of the relevant stellar envelope is 
then determined by balancing the angular momentum gained from the
accretion with that lost from the MHD stellar wind.
We have the following approximate equation
\beq
M_*R_*^2{d\Omega_{*}\over dt}
=-{\dot M}_a\Omega_{*}R_*^2+{\dot M}_{a}\Omega_{k,i}R_i^{2}
-B_{p*}^2 \Omega_{*}R_{*}^3\tau_D.
\label{sd0}
\eeq
where $\Omega_{k,i}$ is the Keplerian angular speed at the disk inner edge,
and we have assumed a constant $R_*$. The last term on the right
hand side is due to the magnetic torque applied by the wind 
($\propto B_{p*}B_{\phi*}$). This equation applies
when $R_i$ corresponds to a radius at which the disk is spinning faster
than
field lines of the star at that radius. Thus it is assumed that the disk
accretion can spin up the stellar convective envelope.
There exists a regime in which the star spins fast enough to disrupt
the disk, and angular momentum of the star is lost to the disk.
We do not consider that regime. Note further that if $R_i$ were greater
than the radius at which the co-rotation speed
exceeded the Keplerian speed of the star, there would be no accretion
(Armitage \& Clarke 1996).

Re-arranging, and expanding the Keplerian term, we have
\beq
{{d\Omega_{*}}\over {dt}}+\Omega_{*}({\dot
M}_a/M_*+B_{p*}^2R_*\tau_D/M_*)
\simeq {(G M_c)^{1/2}{\dot M}_aR_i^{1/2}\over R_*^2M_*}.
\label{sd1}
\eeq
In the simple limit of no disk, only the 1st and 3rd terms
of (\ref{sd1}) would contribute and the solution would be 
\beq
\Omega_{*}=\Omega_{*0}e^{-t/t_m},
\label{nodisc}
\eeq
where we define the magnetic spin-down time scale $t_m$
\beq
t_m \equiv {{M_*} \over {B_{p*}^2R_*\tau_D}}.
\label{tm}
\eeq
This quantity is the spin-down time from
MHD powered rotational energy loss.
For typical parameters, 
\beq
t_{m}\sim 50 \left( {M_{*} \over  M_\odot} \right) 
               \left ( {B_{p*}\over 10 {\rm kG}}  \right)^{-2}
	       \left ( {R_*\over 10^{11} {\rm cm}}  \right)^{-1}
	       \left ( {\tau_D\over 10^{5} {\rm sec}}  \right)^{-1} 
{\rm yr}
\label{mag}
\eeq
As we shall see there are two regimes of interest to pursue in analytic
approximation depending on whether $t/t_{m}$ is large or small.

For the kind of disks considered here, the 
second term in (\ref{sd1}) can be ignored compared with the third.
Also, for the maximum accretion rates we consider $(i.e. \sim
10^{-3}M_\odot
{\rm yr})(t/{\rm 1yr})^{-5/4}$, 
a mass shell of 0.01 $M_\odot$ will
at most have a 30\% gain in mass over the lifetime, because of the
form of the time dependence, so we
can assume $M_*$ is a constant after the 1 year initial system formation period.  The ODE  can then be solved
by standard integrating factor techniques, 
and is in fact analogous to an RL circuit.  We find
\beq
\Omega_{*}=CExp\left[-\int^t{dt' \over t_m}\right]
+Exp\left[-\int^t {dt' \over t_m}\right]
\ts 
\int^tExp\left[\int^t {dt \over t_m}\right]
q(t')dt' 
\label{sd3}
\eeq
\noindent where we define
\beq
q(t) \equiv {(GM_c)^{1/2}{\dot M}_a(t)
R_i^{1/2}(t)\over R_*^2M_*}. 
\eeq
Note again that we do not 
consider the solution before 
$t = 1$yr when the accretion disk is still forming.
For the approximate early time solution, $t/1{\rm yr} < t/t_{m} < 1$,  
we set the exponentials equal to 1.  
The result (after the 1 yr disk formation grace period) is then
\beq
\Omega_{*}(t) \simeq \Omega_{*0} +
\omega_0 \left(1 - \left(\frac{t}{1 {\rm yr}}\right)^{-1/14}\right),
\label{grace}
\eeq
where
\beq
\omega_o = 
{3 \ts 10^{8}(G M_{c})^{1/2}{\dot M}_{a0}
R_{i0}^{1/2}\over R_*^2M_*},
\label{grace3}
\eeq
and 
where the subscript $0$ indicates the quantity at
its initial time $(t=1{\rm yr})$.
We have ignored the time dependence in $R_i$ from $\Omega_{*}$, while 
including the implicit time dependence from ${\dot M}_a$. 
This is a good approximation for this regime 
because $R_i^{1/2}$ depends on $\Omega_{*}^{2/7}{\dot M}_{a}^{-1/7}$.
Note that expression (\ref{grace}) allows us
to define the difference between initially fast ($\Omega_{*0} >
\omega_0$)
and slow ($\Omega_{*0} < \omega_0$) stellar rotation.

When $t/t_{m}> 1$
the exponential in (\ref{sd3}) begins to evolve
quickly compared to the
power law decay of the ${\dot M}_a R_{i}^{1/2}$ 
factor inside the last integral.
Factoring this out and noting
that the product of integrals in this
last term then cancel, we have
\beq
\Omega_{*}(t) \sim q(t)t_m
+\left( K  - q(t)t_m \right) e^{({-t/t_m})},
\label{sd4}
\eeq
where $K=e\Omega_{*}(t_m)+ q(t_m)t_m(1-e)$.
For large times $(t/t_{m}) > 1$, 
the first  term on the right of (\ref{sd4}) dominates and we have 
\beq
\Omega_{*}(t)\simeq q(t)t_m = 
{(G M_{c})^{1/2}{\dot M}_a R_i^{1/2}
\over B_{p*}^2 R_*^3\tau_D}= \alpha_{ss}^{1/7}
{(GM_c)^{3/7}{\dot M}^{6/7}_a (H/R)^{3/7}\over B_{p*}^{12/7}R_*^{15/7}\tau_D}\ ({\rm for}\ {\Omega_* \tau_D <1})
\nn \\
\alpha_{ss}^{1/5}{(GM_c)^{3/5}{\dot M}^{6/5}_a (H/R)^{3/5}\over B_{p*}^{12/5}R_*^{3}\tau_D}\ ({\rm for}\ {\Omega_* \tau_D >1}),
\label{sd5}
\eeq
where the latter represents the case for which 
$R_i$ depends on $\Omega_{*}$ from (\ref{ss7a}).
Note that  the equations of (\ref{sd5}) should be used only when 
it produces a value below the corresponding escape speed.

\subsection{Disk Wind Luminosities}

Plugging (\ref{ss5}) into (\ref{1a}),
we have
\beq
L_{dw}={\dot M}_{dw} V_{dw}^2=\epsilon_d
\alpha_{ss}^{1/2} \int{B}^2_\phi\Omega_{k} R^2dR
\sim \epsilon_d\alpha_{ss}^{1/2}{\dot M}_a (R_i/H_i)GM_c/R_{i},
\label{ss6}
\eeq
where ${\dot M}_{dw}$ indicates the outflow rate from the
disk wind and $V_{dw}$ is the disk wind outflow speed.
We now consider several cases which involve using the 
appropriate value of $R_i$ from section 2.3 in (\ref{ss6}), then using
the appropriate solution for $\Omega_{*}$ from section 2.4 in 
(\ref{ss7a}) for $R_i$.  We will also use (\ref{rrl1}) for ${\dot M_a}$.

We first consider the case of slow stellar rotation  ($\Omega_{*0} <
\omega_0$). The maximally luminous case will have the smallest
disk inner radius.  For a given poloidal field,
the toroidal stellar field will be weakest for slowly rotating stars
as per our discussion above, so in this case the
disk inner radius will be the smallest.
When  $R_i$ 
as calculated in (\ref{ss7a})  satisfies $R_i\le R_*$, we
use $R_i=R_*$ and 
the disk luminosity is given by
\beq
L_{dw} \simeq {\epsilon_d\alpha_{ss}^{1/2}GM_c{\dot M}_a
\over R_*}\left({R_*\over H_i}\right)
\sim
10^{37} { \epsilon_d\alpha_{ss,-2}^{1/2}{\dot M}_{a0,22}
M_{c,1} \over  R_{*,11}} \left({t\over {\rm 1yr}}\right)^{-5/4}\left
({R_*/ H_i}\over 10\right)
{\rm erg/s}.
\label{ss6ab}
\eeq
In this, and the expressions which follow, we scale $\alpha_{ss}$ to $0.01$, 
magnetic fields to $10^4$ G, radii to $10^{11}$ cm,
diffusion times to $10^5$ yr, stellar core masses to $1 ~M_\odot$,
stellar shell masses to $10^{-2} ~M_\odot$,
angular speeds to $10^{-4} ~Hz$ and mass loss rates to 
$10^{22} {\rm g\ s^{-1}}$.  These scalings are
indicated by the subscripts.

We now consider the case of an initially rapidly spinning star. 
The disk and star
are strongly coupled through $R_i$ and we must consider the different
temporal regimes relative to the ratios $t/t_m$ and $\Omega_*\tau_D$.
When $t< t_{m}$ the first term on the right hand side of
 (\ref{grace}) dominates.  Then using (\ref{ss7a}) for $\Omega_*\tau_D>1$, 
we have 
\beq
L_{dw} \simeq 
{\epsilon_d\alpha_{ss}^{3\over 14}(GM_c)^{8\over 7}{\dot M}_a^{9\over 7} \over 
B_{p*}^{4\over 7}\tau_D^{4/7}\Omega_{*}^{4\over 7}R_*^{12\over 7}}\left({R_i\over H_i}\right)^{13/7}
\sim
7 \ts 10^{37}{ \epsilon_d\alpha_{ss,-2}^{3\over 14}M_{c,1}^{8\over 7}
{\dot M}_{a0,22}^{9\over 7}
\over
\left(B_{p*,4}\tau_{D,5}\Omega_{*0,-4}\right)^{4\over 7}R_{*,11}^{12\over 7}}
\left({t\over 1 {\rm yr}}\right)^{-45\over 28} \left({R_i/ H_i \over 10}\right)^{13\over 7}
{\rm erg\over s}.
\label{ss6aba}
\eeq
For $t > t_{m}$, using (\ref{ss7a})
(under the assumption
that $B_{\phi*} \gsim B_{p*}$, or equivalently $\Omega_*\tau_D >1$), 
and (\ref{sd5}) in combination with (\ref{ss6}),  
we have
\beq
L_{dw} \simeq 
{\epsilon_d\alpha_{ss}^{1\over 70}(GM_c)^{4\over 5}{\dot M}_a^{3\over 5}B_{p*}^{4\over 5}}\left({R_i \over H_i}\right)^{11\over 5}
\sim  10^{38} \epsilon_d
\alpha_{ss,-2}^{1 \over 70}  M_{c,1}^{4\over 5}
{\dot M}_{a0,22}^{3\over 5}
B_{p*,4}^{4\over 5}
\left({t\over{\rm 50 yr}}\right)^{-3\over 4} \left({R_i /H_i \over 10 }\right )^{11\over 5}
{\rm erg \over s}
\label{ss6abb}
\eeq
In the limit that $\Omega_{*}\tau_D<1$ or 
$B_{p*}>B_{\phi*}$, from (\ref{ss7a}) 
we obtain instead,
\beq
L_{dw} \sim  
{\epsilon_d\alpha_{ss}^{1\over 14}(GM_c)^{8\over 7}{\dot M}_a^{9\over 7} \over 
B_{p*}^{4\over 7}
R_*^{12\over 7}}\left({R_i \over H_i}\right )^{13\over 7}
\sim 7\ts 10^{35}{\epsilon_d\alpha_{ss,-2}^{1\over 14}M_{c,1}^{8\over 7}
{\dot M}_{a0,22}^{9\over 7}
\over
B_{p*,4}^{4\over 7}R_{*,11}^{12\over 7}}
\left({t\over 50 {\rm yr}}\right)^{-45\over 28} 
\left({R_i / H_i\over 10}\right )^{13\over 7}
{\rm erg\over s}.
\label{ss6abbc}
\eeq
For the characteristic numbers used, (\ref{ss6abb}) does not
apply because a check of $\Omega_*\tau_D$ from (\ref{sd5}) reveals
that $\Omega_*\tau_D < 1$ for $t>t_m$ for our parameter regime. 

The above formalism is  valid
when the magnetic luminosity is $\lsim$
the accretion luminosity $G{\dot M_*} M_c/R_i$.
If the magnetic luminosity exceeds the disk luminosity,
then the disk structure would be  disrupted
and the formalism would have to be revisited.

\subsection{Stellar wind luminosities}

The stellar wind luminosity is given by (\ref{1b}).
We note again the 
hidden dependence of the field on the rotational velocity.
If a dynamo generates the stellar field, then both $B_p$ and
$B_\phi$ can depend on $\Omega_{*}$. For example the
rotation would be required to generate the pseudoscalar helicity 
that sustains a steady field in dynamo models.  
Also, the toroidal field may induce poloidal field
by springing outward. 
More subtle dependences should
be considered in future work, but the important point here
is the recognition of some dependence on the rotation,
and thus on the supply of angular momentum from the disk.
Here we simply assume that the poloidal field of the star
is kept at a constant value either by a dynamo or by 
buoyancy of toroidal field, while the toroidal field 
depends linearly 
on the rotational speed as implied by growth from shear.  In this case, 
equation (\ref{1b}) gives 
\beq
L_{sw}\sim \epsilon_s B_{*p}B_{*\phi}\Omega_{*}R_*^3\sim B_{p*}^2
\Omega_{*}^2\tau_D R_*^3.
\label{ss8}
\eeq
We now consider the same temporal regimes for the stellar wind power
that were examined for the disk wind.
For a rapidly rotating star in the $t < t_{m}$ regime,
the first term on the right hand side of (\ref{grace}) is
dominant and from (\ref{ss8}) we then have,
\beq
L_{sw}\simeq \epsilon_sB_{p*}^2\Omega_{*0}^2\tau_D R_*^3
=10^{38}\epsilon_sB_{p*,4}^2\Omega_{*0,-4}^2\tau_{D,5} R_{*,11}^3\ {\rm erg/s}.
\label{ss12b}
\eeq

For a very slowly rotating star in the $t < t_{m}$ regime,
the second term on the right of (\ref{grace}) is dominant. 
The spin evolves to within a factor of 1/10 of 
$\omega_o$ after a few  years.  In this case $\Omega_{*}$ is
such
that $R_i$ from  (\ref{ss7a})
is less than $R_*$ so we take $R_i=R_*$.
The stellar rotation rate and wind power thus approach 
\beq
\Omega_{*}\sim (3\ts 10^7) \sqrt{GM_c{\dot M}_{a0}^2\over R_*^3M_{*}^2},
\label{grace2}
\eeq
and 
\beq
L_{sw}\sim 3 \ts 10^{31} \epsilon_s {B_{p*,4}^2\tau_{D,5}M_{c,1}
{\dot M}_{a0,22}^2\over M_{*,1}^2}
\ {\rm erg/s}.
\label{ss12b1}
\eeq
respectively. The slow time evolution in (\ref{grace}),
which results in the approximation that the angular speed
and luminosity are constant in (\ref{grace2}) and (\ref{ss12b1}),
reflects the fact that during this period, the angular momentum gain from
accreted material and the loss of angular momentum from the wind 
nearly balance.

For $t>t_{m}$, 
but for $\Omega_{*}\tau_D>1$,  
using  (\ref{sd5}) and  (\ref{ss7a}) 
with (\ref{ss8}), we have
\beq
L_{sw}\sim 
{\epsilon_s\alpha_{ss}^{2\over 5}(GM_c)^{6\over 5}{\dot M}_a^{12\over 5} \over
B_{p*}^{14/5}R_*^3\tau_D}
\left({H_i \over R_i}\right )^{6\over 5}
=10^{28}
{\epsilon_s \alpha_{ss,-2}^{2\over 5}M_{c,1}^{6\over 5}
{\dot M}_{a0,22}^{12\over 5} 
\over B_{p*,4}^{14\over 5}R_{*,11}^3\tau_{D,5}}
\left({t\over {\rm 50 yr}}\right)^{-3}\left({R_i/H_i \over 10}\right )^{-6\over 5}
{\rm {erg \over s}}.
\label{ss12a}
\eeq
For $t>t_{m}$, when $\Omega_{*}\tau_D<1$, we have instead
\beq
L_{sw}\sim 
{\epsilon_s \alpha_{ss}^{2\over 7}(GM_c)^{6\over 7}{\dot M}_a^{12\over 7} \over
B_{p*}^{10\over 7}R_*^{9\over 7}\tau_D}\left({H_i \over R_i}\right )^{6\over 7}
=10^{30}{\epsilon_s \alpha_{ss,-2}^{2/7}M_{c,1}^{6\over 7}
{\dot M}_{a0,22}^{12\over 7} 
\over B_{p*,4}^{10\over 7}R_{*,11}^{9\over 7}\tau_{D,5}}
\left({t\over {\rm 50 yr}}\right)^{-15\over 7}\left({R_i/H_i \over 10}\right )^{-6\over 5}
{\rm {erg\over s}},
\label{ss12ab}
\eeq
which is larger in comparison to (\ref{ss12a}) because
a smaller field means a slower decay of $\Omega_*$. 
For the characteristic numbers used,  
$\Omega_*\tau_D$ from (\ref{sd5}) satisfies
$\Omega_*\tau_D < 1$ for $t>t_m$, as also mentioned above, 
so here only (\ref{ss12ab}) would apply for our parameter regime and not 
(\ref{ss12a}).

\subsection{Slow and Rapid Initial Stellar Rotation Cases}

In fig 1 we show plots of the disk to stellar wind
luminosity ratios using $\epsilon_s =\epsilon_d$ for two different
stellar rotation regimes.

\noindent{\bf Slow Rotation:}
In fig 1a and 1b plots of $L_{dw}/L_{sw}$ for the slow rotator
case ($\Omega_{*0} < \omega_0$) are given. 
Fig 1a shows essentially the ratio of (\ref{ss6ab}) 
to (\ref{ss12b1}) and fig 1b shows
the ratio (\ref{ss6ab}) to (\ref{ss12ab}).
For both the $t<<t_{m}$ and $t>>t_{m}$ regimes,
the disk wind luminosity
strongly dominates the stellar wind luminosity by  orders
of magnitude. This dominance decreases with time in the former
regime and increases in the latter regime.  However the main
feature is that for the entire range, the disk wind dominates
the stellar wind.  

The star is therefore slaved to the disk throughout the
evolution in this case.  The stellar wind and spin axes 
are also slaved to those of the disk because the toroidal field
is stretched perpendicularly to the axis of rotation and the wind
emanates perpendicularly to the toroidal field.  The disk and stellar 
winds would thus be coaxial, but the disk wind would strongly dominate.

\noindent{\bf Rapid Stellar Rotation:} 
In fig 1c and 1d we present plots of $L_{dw}/L_{sw}$ for the rapid rotator
case ($\Omega_{*0} > \omega_0$). 
For $t<t_m$ the plot shows basically the ratio of (\ref{ss6aba}) to 
(\ref{ss12b}) and for $t>t_m$ the plot shows
the ratio of (\ref{ss6abbc}) to
(\ref{ss12ab}).
For $t<t_{m}$, the stellar wind is comparable to, and
even slightly dominates
the disk wind luminosity for the parameters used,
while for $t>t_m$ the disk wind dominates strongly.
As in the slow rotator case, the disk wind domination
decreases slowly for $t<t_m$ and increases for $t>t_m$.



Thus for the large $\Omega_{*0}$ case, with $\Omega_{*0}$ near
maximal speeds, the stellar wind is independent
of the disk until $t=t_{m}$. After this time its MHD wind luminosity
is negligible compared to that of the disk.  (If there were no disk at
all, one would see only the rapid fall of of the stellar  wind as in
equation (\ref{nodisc})).  The initial independence of the star from the
disk means that the axes of the two winds can be misaligned.  The origin
of the stellar spin and magnetic axes are decoupled since the spin
axis of the disk is independently determined by the initial plane of
the binary orbit. Thus the disk and stellar wind system can produce
a multi-polar wind system of up to $\sim 10^{38}$erg/s, depending
on the efficiency of conversion of rotational energy.

\section{Discussion}

Our results show that powerful magneto-centrifugal winds can be driven
in PNe systems. In the context of accretion disks
formed from the break-up of a secondary star in a binary system,
these winds are  time dependent and have a finite lifetime.  Thus
our results would predict that transient, magnetized  outflows can be
driven during the transition from AGB to PNe.  
Magneto-centrifugally driven winds potentially solve two
problems which have emerged from PNe studies (Frank 2000). 
First, the large powers computed above can explain how strong winds 
can be driven independently of radiation line driving processes
as required by observations in the PPNe phase (e.g. Alcolea et al 2000).  
Second, they may explain the high degree of collimation seen in some
sources (e.g. Sahai \& Trauger 1998), and 
multi-polar outflows when the disk and stellar winds are contemporaneous.
Below we discuss these points further. 


\subsection{\bf Accounting for the power of PN or PPN winds:}
The calculations above show that magnetically driven winds
can account for the large powers observed in PPNe.
Alcolea et al. (2000) make the case that observations of PPNe are 
the best place to infer the required powers
and shaping mechanisms of PNe.  None of the 3 objects for which  
Alcolea et al. (2000) have measured kinetic ages  
(OH 231.8; M 1-92; M2-56) can be radiation driven at the PPNe stage. 
The required outflow powers range from $10^{35}$ erg/s
to $10 ^{37}$erg/s.  The kinetic ages are fairly similar, ranging
from 750 yr to 1800 yr. The radiation driving
falls short of being able to provide the power by at least 
two orders of magnitude.  The 9 objects for which Alcolea 
et. al (2000) have measured wind energies but not kinetic ages 
look reasonably  similar in terms of radiation power and energetics.  
Since the kinetic 
ages for the few measured objects are not widely varying, it is likely
that others in the population of 9 also probably require high outflow powers.

The powers computed in section 2 for MHD wind driving 
can be as high as $10^{38}$ erg/s for the parameters used.
Such powers are therefore sufficient for PNe and PPNe.
There is likely to be some efficiency factor with which the driving can take
place, so to explain a $10^{37}$ erg/s wind, we would require an
efficiency factor of $\epsilon \sim 10\%$.  More study
is needed to determine the $\epsilon$ factors dynamically.

For $t> t_m$, the disk wind power falls by 
two orders of magnitude or so, while the stellar MHD wind is of negligible
power.  The MHD disk wind could still supply e.g. 
$\sim 10^{36}$ ergs/s after 
$\sim 50$ yr, continuing to fall as $t^{1.6}$ from (\ref{ss6abbc}). 
Specific examples of kinetic powers required in this approximate range
include Hb5 ($V_{w} \lsim 400$ km/s 
and $\dot M \sim 6\ts 10^{-5}{\dot M}_\odot$/yr; [Corradi \& Schwarz  1995;
Pishmish et al. 2000])
thus with a kinetic luminosity $L_w\lsim 3.6\ts 10^{36}$erg/s; 
IC4406 ($V_{w}\sim 65$ km/s [Corradi al. 1997]
and $\dot M\sim 10^{-4}{\dot M}_\odot$/yr [Sahai et al. 1991; Cox et al. 1991])
with $L_w \sim  2 \ts 10^{35}$erg/s; 
M2-9 ($V_{w}\sim 46$ km/s [Solf 2000]
and $\dot M\sim 4 \ts 10^{-6}{\dot M}_\odot$/yr [Zweigle et al. 1997])
with $L_w \sim 5 \ts 10^{33}$erg/s.
Observations of NGC 3242,
NGC 6828, NGC 7009, and NGC 7662  PNe 
(Balick et al. 1998; Dwarkadas \& Balick 1998) are also
consistent with $10^{33}{\rm erg/s}\le L_w\le 10^{35}{\rm erg/s}$. 

In all cases, a dynamo in either a disk or the star or both may be responsible
for the field driving the wind.  Magnetic flaring and the production of hard
coronal X-rays might then be expected in PPN engines.
Observations with Chandra and other X-ray telescopes are 
therefore be highly desirable.

\subsection{\bf Accounting for the shapes of PN or PPN winds ?:}
Given that the fields can power jets, they may also play 
a role in shaping them. One possible approach to MHD collimation
in these sources is simulated in Garc\'ia-Segura et al. (1999).
Since we have not yet dynamically calculated or 
simulated collimation mechanisms as applied to planetary nebulae
the discussion that follows is speculative.  
The fact that the power of the disk and stellar winds can be comparable 
motivates us to proceed.

It is known that in the context of disks (e.g. K\"onigl \& Pudritz 2000 for review) and even in the context of stars 
(Tsinganos \& Bogovalov 2000) 
that  magnetic self collimation can be 
an important effect. The literature on collimation of
disk winds is extensive. It is noteworthy that
in emphasizing that fast rotating stellar magnetospheres 
can also collimate winds, Tsinganos \& Bogovalov (2000) point out that
such collimation from an underlying rotator 
will likely be effective when the quantity
\beq
Q=0.12
\left({\psi \over\psi_\odot}\right)\left({\Omega\over\Omega_\odot}\right)
\left({ {\dot M}_{w} \over {\dot M}_\odot}\right)^{-1/2}
\left({V_{w} \over V_\odot}\right)^{-3/2}
\label{colcond}
\eeq 
is greater than 1.  Here $\psi$ is the magnetic flux  anchored
on the rotator spinning at angular speed $\Omega$, 
${\dot M}_{w}$ is the outflow mass loss rate, $V_{w}$ is the outflow
speed, and these values are scaled to the respective values for the Sun,
namely $V_\odot=400$km/s, ${\dot M}_\odot=1.6\ts 10^{12}$g/s, $\Omega_\odot
=3\ts 10^{-6}$/s and $\psi_\odot=10^{22}$G-cm$^2$.
We then have 
\beq
Q/Q_\odot \simeq 4.2 \left({B_{*,4}R_{*,11}^2}\over 10^{26}{\rm G\cdot cm^2}\right)\left({\Omega_*\over10^{-4}{\rm /s}}\right)
\left({ {\dot M}_{sw} \over 6\ts 10^{21}{\rm g/s}}\right)^{-1/2}
\left({V_{sw} \over 400{\rm km/s}}\right)^{-3/2},
\label{colcond1}
\eeq 
where we have scaled to our fiducial post-AGB star parameters, taking the 
upper limit of the magnetic flux, and
assuming a kinetic luminosity of the wind 
$L_{kin}\simeq {\dot M}_{sw}V_{sw}^2 \sim 10^{37}$erg/s.
We can see that $Q/Q_\odot$ can be greater than 1 
for the chosen parameters, and
thus collimation in excess of that for the solar wind can occur.  
If we instead use representative outflow parameters for later
stages of PNe (e.g. Balick et al. 1998), 
such as 
$\dot{M} = 5\times 10^{-7} M_\odot \: {\rm yr^{-1}}$ and $V = 200 \: \rm{km \: s^{-1}}$), 
we find $Q/Q_\odot\gsim 100$ which implies significant collimation.
In general, these estimates suggest that 
self-collimation of magnetized winds in PPNe and PNe
might be possible without requiring shocks to aid in the process.
The shock-aided approach has been studied in 
Garc\'ia-Segura et al. (1999). More study is warranted.
The relative self-collimation of the disk wind and the stellar wind
may be different for different systems. This needs to be studied.

Assuming collimation can ensue, 
the shape of any multi-polar magnetically driven outflow in our picture
will also depend on the stellar spin down time scale $t_m$, and 
the time scale for the star to become a PN, $\equiv t_{PN}$.
The ratio of $t_m/t_{PN}$ will determine how much 
shaping will be controlled by the a combined star+disk wind.  When
$t_m/t_{PN} << 1$, the stellar wind is likely to provide 
micro-structures such as knots, bullets, or ansae in mature PNe.  When 
$t_m/t_{PN} \gsim 1$ then structures such as multi-polar bubbles, which
develop during the period when $L_{dw} \approx L_{sw}$, may leave a 
large enough imprint to survive into the PN phase. 

Another important parameter is the ratio of the
wind momentum densities, $p_{dw}/p_{sw}={\rho}_{dw} V_{dw}/
({\rho}_{sw}V_{sw})$ where the ratio to be used here 
is that computed for each type of 
wind independent of the actual presence of the other.  
The momentum density depends on collimation and is thus a function of angle
from a wind's axis. 
The terminal wind speed of material at a given radius from any wind axis  
depends on the details of the wind launching
process but should be of order several times the 
the escape speed at the radius at which the field lines
threading that radius are anchored 
(e.g. K\"onigl \& Pudritz 2000 for a review).  
The ratio $p_{dw}/p_{sw}$ can in part
determine the shape of the resulting observed outflow combination  
by determining which wind-driven 
``lobe'' would extend farther when the two winds are superimposed.  
The difference in initial launch times of 
the two winds is also very important in determining the observed
shapes.

For two concurrent winds satisfying $p_{dw}/p_{sw}  > 1$ on average,
 and for which the star is rotating
slowly ($\Omega_{*o} < \omega_o$), we might 
expect two nested aligned bipolar bubbles (as shown in Fig 2a)
when the terminal speed of the stellar wind, 
calculated independently from the disk
wind, is faster over most of the outflow cross-section. 
The stellar wind can sweep up disk material, but because of the lower overall 
momentum density, it cannot proceed as far in distance. 
An example of a PN with this shape is Hubble 12 (Welch \ea 1999).  
When the stellar and disk magnetic and rotational
axes are misaligned, a nested multi-polar bubble may emerge as 
shown in Fig 2c.  A possible 
example of such an outflow is M2-46 (Manchado 2000)
Please note again: Fig 2 represents cartoon speculations, not simulations.

When $p_{dw}/p_{sw} < 1$ and the star is rapidly rotating
($\Omega_{*o} > \omega_o$),  the stellar wind could push ahead of
the disk wind and may produce the features seen in e.g. NGC 7007
(Fig 2b, Balick \ea 1998).  In the case where the star is initially
rapidly rotating 
and the stellar and disk axes are not aligned, then a second form
of the nested multi-polar bubble may emerge as is shown in Fig 2d.
An example of such an outflow may be He2-155 (Sahai 2000).

In the rapidly rotating case,
the stellar wind remains more or less constant over a spin-down period
while the disk wind falls.  Thus we might expect the stellar wind
to appear more or less continuous, whereas the disk wind
could produce structures only of comparable luminosity to the stellar
wind for the first several years.   In this initial phase, 
antipodal knots amidst a continuous stellar wind might appear
where the disk and stellar winds interact, or 
where the disk interacts with the ambient medium.
Such processes might have been occurring in the Egg nebula (Jura et al. 2000).  

Given that typical sound speeds for observed outflows are of order
20-30 km/s, while the wind velocities can be $>100$km/s, we can
estimate the size of outflow structures compared with the distance
from the star.  For a hybrid disk-star 
outflow with strongest power phase lasting
$\sim$ few years, but with an age or order hundreds of years, one
should see knots
(or shells depending on collimation) with a width of order
$10^{16}$ cm. If the outflow velocity were 100 km/s, then
the knot/shell distance from the central core would be $\sim 5$ times
the shell thickness.  Such features are seen in the Cat's Eye nebula
(Miranda \& Solf 1992)

Note that two interacting winds can lead to more than 2 multiple
symmetry axis due to edge features.
This and the features described above could also result from 
time dependent effects of separate interacting
stellar winds or separate interacting disk winds where the star or disk
is allowed to precess (e.g. Manchado et al. 1996).  
Precession would enable successive outbursts 
to be emitted on multiple axes.  If there is precession, then one might see 
an additional point symmetry.
These ideas may  be applied to other coupled disk+star systems
(Soker \& Livio 1994, Mastrodemos \& Morris 1998).  

\section{Summary and Conclusions}

We have shown that MHD winds from either disks or post-AGB
stars can comfortably power PNe outflows using reasonable
estimates of stellar and disk field strengths and rotation
parameters.  The available MHD wind power 
turns out to be many orders of magnitude larger than that
available for radiation/line driven outflows inferred from observations
of  PPNe  (e.g Alcolea et al. 2000).
Our study also highlights the importance of understanding
the interplay between disk winds and stellar winds in contributing to
the diversity of observed multi-polar PNe morphologies.


We derived maximal disk and stellar MHD wind powers for a coupled
disk-star system at times $t<t_{m}$, $t>t_{m}$, (where $t_m$ is the stellar
spin down time) and for 
fast and slow stellar rotators, $\Omega_*/\omega_0 < {\rm or} > 1$,
and values of  $\Omega_*\tau_D < {\rm or} > 1$.
Our calculations take the initial time to be the onset of disk formation in
a binary system from common envelope expulsion.  The initial
onset of the star and disk wind should be nearly contemporaneous. The
layers of the star with the strong magnetic field that power the
wind are only exposed after the common envelope is ejected, and the
accretion disk forms within  a year from the common envelope ejection.

The resulting luminosities for both winds depend only on
the disk accretion rate and properties of the star.  This is because
the inner regions of the disk are the main contributor to the disk
wind and the disk properties can be  
constrained using properties of magneto-shearing 
disks and the interaction with the star.  
\ni We now summarize some numbers for the maximum wind powers:  

\ni {\bf *} For $t<<t_m$ and our choice of fiducial scalings, 
the maximum MHD luminosity of a stellar wind from an initially rapidly rotating
post AGB star with field strengths consistent with those estimated for AGB
dynamos 
of $\sim 10^4$ Gauss 
satisfies $L_{sw}\lsim 10^{38}$ erg/s, and is 
approximately constant in time. The disk wind satisfies
$L_{dw}\lsim 7\ts 10^{37}(t/{\rm 1 yr})^{-1.6}$ as the accretion rate falls.

\ni {\bf *} After  $t_{m}>> 50 {\rm yr}$, $L_{sw} \lsim 10^{30}(t/50{\rm
yr})^{2.1}$, and the disk wind power goes  as $L_{dw}\lsim
7\ts 10^{35}(t/{\rm 50yr})^{-1.6}$.  
 
\ni {\bf *} 
Whilst the dominance of disk vs. stellar wind depends
on $\Omega_{*0}$ for $t<< t_{m}$, 
for all $t>>t_{m}$, the disk wind always dominates, regardless of 
$\Omega_{*0}$.

Since the MHD disk winds dominate the stellar winds for
$t>t_{m}$ regardless of the initial stellar spin rate, 
the star is slaved to the disk in that regime.
For $t< t_{m}$, the stellar wind power can equal or exceed 
that of the disk,  and in this regime 
the stellar wind can emanate along a different symmetry axis
than the disk wind.  Multi-polar outflows can then be produced.
For a rapidly rotating star, the stellar MHD wind power is steady for
a time of order $t=t_{m}$, while
the disk wind power  falls rapidly.  For $t> t_{m}$,
the disk wind power falls less rapidly than the stellar wind power.  

Generally, we would suggest that 
if a sustained MHD driven bipolar outflow extending to the core
lasts for $>500$ yr, the disk is dominant. The 
choice of scalings we use in the text were mainly to provide a general 
framework. The spin down time $t_m$ could be longer than our scale of 50 yr
depending on radius and stellar convective diffusion time, (c.f. eqn (\ref{tm})). Disk winds imply initial formation from a binary system.
An initially rapidly 
rotating star would slow down rapidly by this later time, and would only be
able to contribute significantly if its wind were radiatively driven at this
stage. If a stellar wind operates powerfully during $t<t_{m}$ it means that
the central star was initially rotating rapidly.  

There is no guarantee that a binary system would form, or that the star
would be initially rapidly rotating 
for every post AGB star.  Thus for all systems
with bipolar outflows, observations that can reveal the rotation rate
of the central star, its magnetic field, or the presence of binary
companions are ultimately essential for testing MHD disk/stellar
wind paradigms.
Further studies of the disk wind-stellar wind interplay are most
certainly needed.  Given the ubiquity of the accretion-central
object-outflow connection throughout the universe the same kind of
investigation may ultimately be important for other stellar and compact
object accretion systems as well.  

\ni {\bf Acknowledgements}: E.B. was supported in part by 
NSF grant PHY94-07194 at the ITP, and acknowledges 
support from a DOE Plasma Physics Junior Faculty Development Award.  
E.B. also thanks R. Krasnolpolsky,
R. Blandford, B. Chandran, and E. Ostriker for discussions.
Many thanks to G. Garc\'ia-Segura and M. Reyes-Ruiz
for their careful readings which led to important changes.

\vfill
\eject
\centerline{\bf Figure Captions}
\medskip
\ni{\bf Figure 1}: Plot of the MHD wind power ratio for the disk and
star for four regimes with $\epsilon_s =\epsilon_d$: 
a) $t< t_m$ for the initially slowly rotating
star
b) $t >t_m$ for the initially slowly spinning star  
c) $t< t_m$ regime for initially fast rotating star. d) $t> t_m$ regime
for initially fast rotating star. See text.
\medskip

\ni{\bf Figure 2}: Cartoon representing our speculation on
a possible correspondence between between HST PN images and
multi-polar wind structures from the hybrid stellar-disk
wind paradigm, for different initial stellar spins and thrust ratios
$\Pi_{dw}/\Pi_{sw}$ (see text).  (Note that these are not
simulations, but speculations.)
a) top left: $\Omega_{*0}< \omega_0$
and  $\Pi_{dw}/\Pi_{sw} > 1$ so star is slaved to disk and the only
possibility is nested winds. Inset is PN Hubble 12 (Welch \ea 1999).
b) top right: $\Omega_{*0}<\omega_0$ and $\Pi_{dw}/\Pi_{sw} < 1$.
Inset is NGC 7007 (Balick \ea 1998).  c) bottom left: $\Omega_{*0}>
\omega_0$ and $\Pi_{dw}/\Pi_{sw} > 1$.  Inset is M2-46 (Manchado \ea
2000).  d) bottom right: $\Omega_{*0}> \omega_0$ and $\Pi_{dw}/\Pi_{sw}
< 1$.  Inset is He2-155 (Sahai 2000). See text.

\vfill
\eject

\centerline {\bf References}

\ni Alcolea J., Bujarrabal V., Castro-Carrizo A., 
S\'anchez Contreras C., Neri R., Zweigle J. 2000, in
{\it Asymmetrical Planetary Nebulae II: From Origins to Microstructures}, 
ASP Conference Series, Vol. 
199. Edited by J. H. Kastner, N. Soker, and S. Rappaport, p347.

\ni Armitage P.J. \& Clarke C.J., 1996, MNRAS, 280, 458.

\ni Balick, B. 1987,  AJ, 94, 671

\ni Balick B., in {\it Asymmetrical Planetary Nebuale II}, 2000
J.H. Kastner, N.Soker, \& S. Rappaport eds., ASP Conf. Ser. Vol. 199.,
pg 41

\ni Balick, B; Alexander, J., Hajian, A., Terzian, Y,
Perinotto, M., Patriarchi, P., 1998, AJ, 116, 360

\ni Balbus S.A. \& Hawley J.F., 1991, ApJ , 376 214.

\ni Balbus S.A. \& Hawley J.F., 1998, Rev Mod Physics, 72 1.

\ni Blackman E.G., Frank A., Markiel A., Thomas J., Van Horn H.M., 
2000, submitted to Nature.

\ni Blackman E.G., Yi I., Field G.B., 1996, ApJ, 473, L79.

\ni Blackman E.G., 2000, ApJ 529, 138.

\ni Blandford R.D. \& Payne D.G. 1982, MNRAS, 199 883.

\ni Blandford R.D., 2000, in 
``Proc of Discusison Meeting on Magnetic Activity in Stars, Discs and Quasars.'', Ed. D. Lynden-Bell, E. R.
Priest and N. O. Weiss. To appear in Phil. Trans. Roy. Soc. A

\ni Chevalier, R., \& Luo, D., 1994, ApJ, 421, 225

\ni Contopoulos J., 1995, ApJ 450 616.

\ni Corradi R.L.M. \& Schwarz H.E., 1995, A\&A 
293 871.

\ni Corradi R.L.M., Perinotto M., Schwarz H.E., 
Klaeskins J.F., 1997, A\&A  322 975.

\ni Cox, P., Huggins P. J., Bachiller R.,
 Forveille T., 1991, A\&A 250, 533.

\ni Duncan R.C. \& Thomson C., 1992, 392 L9

\ni Dwarkadas V.V. \& Balick B., ApJ, 1998, 497 267.

\ni Ferrari A., 1998, ARAA, 36 539.

\ni Frank A., 1999, New Ast. Rev., 43, p31.

\ni Frank, A., in {\it Asymmetrical Planetary Nebuale II}, 2000
J.H. Kastner, N.Soker, \& S. Rappaport eds., ASP Conf. Ser. Vol. 199.,
pg 225


\ni Garc\'ia-Segura, G., 1997, ApJ, 489L, 189

\ni Garc\'ia-Segura G., Langer N., Rozyczka, M., Franco J, 1999,
ApJ 517 767.

\ni Ghosh P.\& Lamb F.K., 1978, ApJ 223, L83. 

\ni Iben I., 1991, ApJS 76, 55.


\ni Icke, V., 1988, A\&A, 202, 177

\ni Jura M., Turner J. L., Van Dyk S., \& Knapp G.R., 2000,
ApJ, 528 L105.

\ni Kahn, F. D., \& West, K. A., 1985, MNRAS, 212, 837

\ni K\"onigl, A., \&  Ruden, S.P. 1993, in ''Protostars and Planets
III'', ed.
E.H. Levy \& J.I. Lunine (University of Arizona Press), 641.

\ni K\"onigl, A.; Pudritz, R. E., 2000
in {\sl Protostars and Planets IV},
 eds Mannings, V., Boss, A.P., Russell, S. S.), 
(Tucson: University of Arizona Press; p. 759

\ni Kwok, S., Purton, C., Fitzgerald, P. M., 1978, ApJ 219, L125

\ni Lovelace R.V.E., Wang J.C.L., Sulkanen M.E., 1987  ApJ, 315 504.


\ni Lynden-Bell D., 1996, MNRAS, 279 389.

\ni Manchado A., Stanghellini L, Guerrero M.A., 1996, ApJ, 466 L95.

\ni Manchado A., Villaver, E., Stranghelli, L., Guerrero, M.,
in {\it Asymmetrical Planetary Nebuale II}, 2000
J.H. Kastner, N.Soker, \& S. Rappaport eds., ASP Conf. Ser. Vol. 199, pg
17 

\ni Mastrodemos, N., Morris, M.,  1998, ApJ, 497, 303

\ni Markiel J.A. \& Thomas J.H., 1999, ApJ, 523 827.

\ni M\'esz\'aros, P., \& Rees, M.J. 1997, Ap.J. 482, L29

\ni Miranda L.F. \& Solf J., 1992, A\&A 260 397.

\ni Mirabel I.F. \& Rodriguez L.F., 1999, ARAA, 37 409.

\ni Morris, M. 1987, PASP, 99, 1115

\ni Ostriker E.C. \& Shu F.H., 1995, ApJ, 447, 813

\ni Osterbrock, D., 1989, {\it Astrophysics of Gaseous Nebulae and
Active
Galactic Nuclei}, (University Science Books, CA)

\ni Parker E.N., {\sl Cosmical Magnetic Fields} 
(Oxford Univ Press: Oxford) 1979.
 
\ni Parker E.N., 1993, ApJ, 408, 707.

\ni Pascoli G., 1997, ApJ, 489, 946.

\ni Pishmish P., Manteiga M., Mampaso Recio A., 2000
in {\it Asymmetrical Planetary Nebuale II}, 2000
J.H. Kastner, N.Soker, \& S. Rappaport eds., ASP Conf. Ser. Vol. 199, pg
397.  

\ni Pudritz, R.E. 1991, in ``The Physics of Star Formation and Early
Stellar
Evolution'', eds. C.J. Lada and N.D. Kylafis, NATO ASI Series (Kluwer),
pg 365.

\ni Pelletier G. \& Pudritz R.E., 1992, ApJ, 394 117.

\ni Reyes-Ruiz M. \& Lopez J.A., 1999, ApJ, 524 952.

\ni Reyes-Ruiz M. \& Stepinksi T.F., 1995, ApJ, 438 750.

\ni Rozyczka, M., \& Franco, J., 1996, ApJ, 469, 127

\ni Ruderman M., Tao L., Kluzniak W., 2000, sub to ApJ, astro-ph/0003462

\ni Sahai R., Wootten A.,
 Schwarz H.E., Clegg R. E. S, 1991, A\&A 251 560.

\ni Sahai, R., \& Trauger, J. T. 1998, AJ, 116, 1357

\ni Sahai, Raghvendra; Trauger, John T., Watson, Alan M., Stapelfeldt,
Karl R.,
Hester, J. J., Burrows, C. J., Ballister, G. E., Clarke, J. T., Crisp,
D., Evans, R. W., Gallagher, J. S., III;
Griffiths, R. E., Hoessel, J. G., Holtzman, J. A., Mould, J. R.,
Scowen, P. A., Westphal, J. A.,  1998, ApJ, 493, 301

\ni Sahai, R., in {\it Asymmetrical Planetary Nebuale II}, 2000
J.H. Kastner, N.Soker, \& S. Rappaport eds., ASP Conf. Ser. Vol. 199, pg
209.

\ni Sch\"onberner D., 1993, in IAU Symp 155, {\sl Planetary Nebulae}
ed. R. Weinberger \& A. Acker (Dordrecht: Kluwer) 415.

\ni Shakura N.I. \& Sunyaev R.A., 1973, A\&A 24 337. 

\ni Shu F.H., Najita J., Ostriker E., Wilkin F., Ruden S.,
Lizano S., 1994, ApJ 429 781.

\ni Smith M.D., 1998, Ap\&SS 261 169.

\ni Soker, N., Livio, M., 1994, ApJ, 421, 219

\ni Soker N., 1998, MNRAS, 299, 1242.

\ni Tsinganos, K., \& Bogovalov
S., 2000,
Magnetic Collimation of Solar and Stellar Winds, {\it Astron.
Astrophys.} in press.

\ni Uchida Y. \& Shibata K. 1985, PASJ, 37 31.

\ni Usov V.V., 1992, Nature, 357, 472

\ni Welch, C., Frank, A., Pipher, J., Forrest, W., Woodward, C., 
1999, ApJ, 522L, 69

\ni Zweigle J., Neri R., Bachiller R.,
 Bujarrabal V., Grewing M., 1997, A\&A, 324 624.

\end{document}